\documentclass[10pt,twocolumn,letterpaper]{article}

\usepackage[pagenumbers]{wacv} 

\usepackage{graphicx}
\usepackage{amsmath}
\usepackage{amssymb}
\usepackage{booktabs}

\usepackage{xcolor}         
\usepackage{algorithm}
\usepackage{algpseudocode}

\usepackage{siunitx}  

\usepackage{multirow}
\DefineNamedColor{named}{PineGreen}     {cmyk}{.92,0,0.59,0.25}
\usepackage{mathabx}
\usepackage{txfonts}
\usepackage{arydshln}

\usepackage[pagebackref,breaklinks,colorlinks]{hyperref}

\usepackage[capitalize]{cleveref}
\crefname{section}{Sec.}{Secs.}
\Crefname{section}{Section}{Sections}
\Crefname{table}{Table}{Tables}
\crefname{table}{Tab.}{Tabs.}

\def\wacvPaperID{158} 
\def\confName{WACV}
\def\confYear{2025}

\begin{document}

\title{MulModSeg: Enhancing Unpaired Multi-Modal Medical Image Segmentation with Modality-Conditioned Text Embedding and Alternating Training}

\author{
Chengyin Li$^{1,2}$ \quad Hui Zhu$^{1}$ \quad Rafi Ibn Sultan$^{1}$ \quad Hassan Bagher Ebadian$^{2}$ \quad \\ Prashant Khanduri$^{1}$ \quad Chetty Indrin$^{3}$ \quad Kundan Thind$^{2}$ \quad Dongxiao Zhu$^{1}$\thanks{Corresponding author.}\\ 
 $^{1}$Wayne State University \quad $^{2}$Henry Ford Health \quad $^{3}$Cedars Sinai Medical Center\\
{\tt \small \{cyli, hui, rafisultan, khanduri.prashant, dzhu\}@wayne.edu} \\ {\tt \small \{cli6, hbagher1, kthind1\}@hfhs.org \quad indrin.chetty@cshs.org}
}

\maketitle

\begin{abstract}
In the diverse field of medical imaging, automatic segmentation has numerous applications and must handle a wide variety of input domains, such as different types of Computed Tomography (CT) scans and Magnetic Resonance (MR) images. This heterogeneity challenges automatic segmentation algorithms to maintain consistent performance across different modalities due to the requirement for spatially aligned and paired images. Typically, segmentation models are trained using a single modality, which limits their ability to generalize to other types of input data without employing transfer learning techniques. Additionally, leveraging complementary information from different modalities to enhance segmentation precision often necessitates substantial modifications to popular encoder-decoder designs, such as introducing multiple branched encoding or decoding paths for each modality. In this work, we propose a simple Multi-Modal Segmentation (MulModSeg) strategy to enhance medical image segmentation across multiple modalities, specifically CT and MR. It incorporates two key designs: a modality-conditioned text embedding framework via a frozen text encoder that adds modality awareness to existing segmentation frameworks without significant structural modifications or computational overhead, and an alternating training procedure that facilitates the integration of essential features from unpaired CT and MR inputs. Through extensive experiments with both Fully Convolutional Network and Transformer-based backbones, MulModSeg consistently outperforms previous methods in segmenting abdominal multi-organ and cardiac substructures for both CT and MR modalities. The code is available in this 
{\href{https://github.com/ChengyinLee/MulModSeg_2024}{link}}.
\end{abstract}

\section{Introduction}
\label{sec:intro}

Medical image segmentation leverages multiple imaging techniques, such as Computed Tomography (CT) and Magnetic Resonance (MR) Imaging, to provide comprehensive views of tissues or organs for disease diagnosis and surgical planning~\cite{li2020towards}. Different modalities offer unique advantages; MR provides superior soft tissue contrast, while CT delivers better bone detail and higher spatial resolution~\cite{metzner2015direct}. Recent advancements in convolutional~\cite{ronneberger2015u,malhotra2022deep} and transformer-based neural networks~\cite{hatamizadeh2022unetr,hatamizadeh2021swin,li2023focalunetr,li2023autoprosam} have achieved competitive segmentation performance. However, while humans can easily identify features across modalities, algorithms trained on a single modality struggle with segmenting multiple modalities. This leads to performance inconsistencies when tested on different image types due to data variability, stemming from factors such as varying imaging methods, scanners, acquisition settings, or patient conditions~\cite{zhou2021review}. Training separate models for each modality would be straightforward, but it would require a massive amount of annotated data and could fail to leverage inter-domain information.

To address this issue, researchers proposed several multi-modal medical image segmentation methods \cite{zhou2019review,zhang2021modality,marinov2023mirror,li2020towards,zhou2022generalizable}.  The first group of methods aims to produce better segmentation by simultaneously utilizing information from multiple modalities. In this context, techniques such as input/layer/decision-level fusion~\cite{zhou2019review}, modality-specific representation~\cite{zhou2021latent}, and hyperdense connections~\cite{dolz2018hyperdense} have been employed to enhance segmentation by integrating information from diverse sources more effectively than single-modality methods. Typically, different modalities are treated as separate inputs for the model, which then generates combined inputs or learns a common representation. However, these methods often require spatially aligned, paired images from the same patient, a condition rarely met due to misalignments and variations in unpaired images, thus compromising performance~\cite{dou2020unpaired}.

Recent advancements in multi-modal learning for medical image segmentation from unpaired CT and MR scans have leveraged the unique attributes of each modality without relying on paired datasets. Research by Dou \etal~\cite{dou2020unpaired} and Jiang \etal~\cite{jiang2021unpaired} introduced efficient architectures that share convolutional kernels between modalities and incorporate modality-specific normalization and innovative loss functions, improving segmentation accuracy. Dual-stream architectures~\cite{elghazy2021multi}, attention mechanisms~\cite{yang2023toward}, and adversarial training~\cite{mondal2018few} have further enhanced performance. However, challenges remain in handling domain shifts between MR and CT images, and methods like the X-shaped architecture~\cite{valindria2018multi} add overhead. Synthetic image generation and semi-supervised learning approaches~\cite{chartsias2017adversarial, liu2023modality} offer alternatives but require significant framework modifications.

In this study, we introduce a versatile Multi-Modal Segmentation (MulModSeg) strategy that seamlessly integrates modality-conditioned text embeddings into any encoder-decoder architecture to enhance multi-modality medical image segmentation without significant architectural modifications during supervised training. Our framework introduces two key innovations: (1) a modality-specific text embedding via the frozen CLIP~\cite{conneau2019cross} text encoder, which brings modality awareness to existing segmentation frameworks, and (2) an alternating training algorithm (ALT) that facilitates the integration of essential features from unpaired images. Inspired by the CLIP-driven universal model concept~\cite{liu2023clip}, MulModSeg self-adjusts its decoder embedding layers to generate precise segmentation outputs, expanding its application to both CT and MR modalities. The ALT strategy enables efficient, one-pass end-to-end supervised training by managing batched CT and MR samples sequentially, allowing a single encoder-decoder structure to accurately segment images across both modalities. Furthermore, our approach can be easily integrated into popular Fully Convolutional Network (FCN)-based and Transformer-based segmentation networks such as UNet~\cite{ronneberger2015u} and SwinUNETR~\cite{hatamizadeh2021swin}.

Our contributions are threefold:
\begin{itemize}
    \item We propose MulModSeg, a method that integrates modality-conditioned text embeddings into segmentation pipelines without requiring significant architectural modification or computational overhead.
    \item We develop an alternating training procedure enabling efficient multi-modal training across CT and MR datasets, ensuring robust performance across imaging modalities.
    \item We validate MulModSeg’s effectiveness through superior results on abdominal multi-organ segmentation (AMOS~\cite{ji2022amos}) and cardiac substructure segmentation (MMWHS~\cite{zhuang2019evaluation}) tasks across strong baseline methods with and without textual inputs.
\end{itemize}


\section{Related Works}
\subsection{Multi-Modality Learning in Medical Imaging}
Recent advancements in deep learning architectures have significantly improved image segmentation performance \cite{milletari2016v, chen2021transunet, hatamizadeh2021swin, hatamizadeh2022unetr}. Among these, UNet~\cite{ronneberger2015u} remains widely used as a baseline for developing more advanced models. More recent transformer-based variations, such as TransUNet \cite{chen2021transunet}, UNETR \cite{hatamizadeh2022unetr}, SwinUNETR \cite{hatamizadeh2021swin}, and FocalUNETR~\cite{li2023focalunetr}, have also demonstrated superior performance. In clinical imaging, multi-modality learning has gained traction, leveraging Generative Adversarial Networks (GANs) \cite{dalmaz_resvit_2022, zhang_multi-contrast_2022} and cross-modality segmentation techniques to exploit inter-domain features without needing registered paired images \cite{zhang_cross-task_2022, zheng_cross-modality_2015, valindria_multi-modal_2018}. Researchers have also explored normalization layers for better generalization \cite{segu_batch_2021, zhou_generalizable_2022}, as well as GANs for reducing the appearance gap between modalities \cite{chen_crdoco_2020, zhang_translating_2019}.

Our framework differs by focusing on modality-specific feature extraction using modality-conditioned text embeddings with frozen text encoders. This approach minimizes the need for significant modification to the existing segmentation models and allows for easy alternation between input modalities during training, unlike prior methods that depend heavily on image translation or prior training.

\subsection{Text Assisted Medical Image Segmentation}

Text-assisted medical image segmentation has shown promise in enhancing accuracy and efficiency by integrating textual information with visual data. The TGANet study~\cite{tomar2022tganet} leveraged text-based embeddings to guide segmentation in colonoscopy, adapting to varying polyp sizes and numbers, thereby outperforming traditional image-only methods. Similarly, Zhong \etal~\cite{zhong2023ariadne} demonstrated language-driven segmentation's benefits, improving Dice scores and reducing training data requirements. Liu \etal~\cite{liu2023m} extended this by using frozen language models to stabilize training and enhance latent space representations, showing superior performance across tasks while reducing computational costs. Chen \etal~\cite{chen2023generative} further expanded vision-language pretraining to 3D medical images by generating synthetic text using large language models, addressing the scarcity of paired text descriptions. Despite these advancements, most methods target specific tasks in a single modality. In this paper, we propose using a frozen text encoder to encode modality-specific information into encoder-decoder architectures for multi-modal medical image segmentation.

\begin{figure*}[t]
	\centering
        \includegraphics[width=0.95\textwidth]{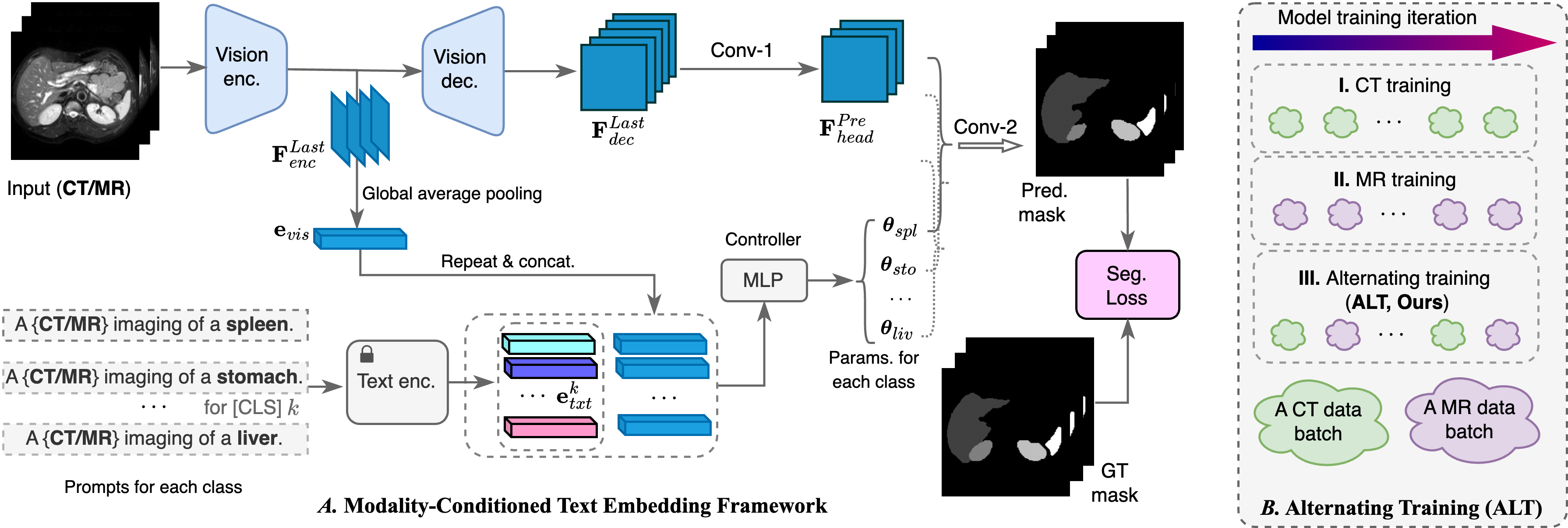}
	\caption{Schematic representation of the MulModSeg strategy for multi-modal medical image segmentation with (A) modality-conditioned text embedding framework and (B) alternating training (ALT).}
 \label{fig: MulModSeg_arch}
\end{figure*}

\section{Methods}
\subsection{Problem Definition}
Consider a set of \(N\) datasets \(\{\mathbb{D}_1^{M_1}, \mathbb{D}_2^{M_2}, \ldots, \mathbb{D}_N^{M_N}\}\), each dataset associated with a specific imaging modality \(M_1, M_2, \ldots, M_N\). Traditionally, separate models are trained for each unpaired dataset, such as \(\mathbb{D}_1^{CT}\) and \(\mathbb{D}_2^{MR}\) in multi-class organ segmentation tasks. This approach requires substantial annotated data and misses the opportunity to leverage complementary information across modalities. To address this, we aim to train a single model capable of handling both \(\mathbb{D}_1^{CT}\) and \(\mathbb{D}_2^{MR}\) in the unpaired scenario.

\subsection{MulModSeg Strategy}
We propose a MulModSeg strategy to tackle the unpaired multi-modal segmentation problem. This strategy comprises two key components: a modality-conditioned text embedding framework (see \Cref{fig: MulModSeg_arch}A) and an alternating training (ALT) method, as illustrated in \Cref{fig: MulModSeg_arch}B. The framework consists of a text embedding branch and a vision branch. In the text branch, a text embedding is generated for each class (or organ) using appropriate medical prompting (see \Cref{tab: eff_text_emb}). The vision branch then takes both CT/MR scans and the text embeddings to predict the segmentation masks. ALT ensures that the model learns iteratively from both CT and MR datasets in a balanced manner during training, thereby addressing the convergence issues associated with mixed-modality at the sample level and eliminating the need to develop separate models for each modality.

\noindent \textbf{Modality-conditioned Text Branch} \label{sec:text_branch}
Let $\mathbf{e}_{txt}^{k}$ denote the text embedding of the $k$-th class (or organ), generated by a pre-trained text encoder such as CLIP, BioBERT, or MedCLIP, combined with a medical prompt (e.g., ``A magnetic resonance imaging of a spleen"). This prompt follows the template ``A \{CT/MR\} imaging of a [CLS]", where [CLS] represents a specific organ (e.g., spleen, stomach, liver), and CT and MR refer to computerized tomography and magnetic resonance, respectively. We first concatenate the text embedding $\mathbf{e}_{txt}^{k}$ with the global image feature $\mathbf{e}_{vis}$ from the vision branch encoder. This concatenated representation is then fed into a multi-layer perceptron (MLP), referred to as the modality-conditioned text-based controller, to generate parameters $\pmb{\theta}_k$, \textit{i.e.}, $\pmb{\theta}_k = \mathrm{MLP}(\mathbf{e}_{txt}^{k} \oplus \mathbf{e}_{vis})$, where $\oplus$ denotes concatenation. Both $\mathbf{e}_{txt}^{k}$ and $\mathbf{e}_{vis}$ are vectors, and $\pmb{\theta}_k$ is also a vector that can be unsqueezed and split into weights and biases~\cite{tian2020conditional,liu2023clip} for the Conv-2 operation (see \cref{fig: MulModSeg_arch}A), which is used to generate the segmentation maps.

\noindent \textbf{Vision Branch} \label{sec:vis_branch}  
The MulModSeg strategy effectively integrates popular U-Net-like architectures, including the FCN-based UNet~\cite{ronneberger2015u} and the transformer-based SwinUNETR~\cite{hatamizadeh2021swin}, into a unified encoder-decoder framework tailored for unpaired multi-modal medical image segmentation. The encoder-decoder backbone follows a ``U" shape, consisting of a multi-staged contracting path for downsampling and a multi-staged expansive path for upsampling, with skip connections between them. The contracting path distills contextual information into a compact form while reducing spatial dimensions, and the expansive path, using skip connections, merges features from the contracting path. This process aids in precise localization and the recovery of spatial details lost during downsampling. Further details on the backbones can be found in the implementation section.

For a 3D volume input $\mathbf{X} \in \mathbb{R}^{1 \times H \times W \times D}$, 1 represents the channel number (for both CT and MR), and $H$, $W$, and $D$ are the spatial dimensions of height, width, and depth, respectively (e.g., $H=W=D=96$). We utilize either the 3D UNet or SwinUNETR backbone to process $\mathbf{X}$ and extract two key feature maps: one from the last stage of the encoder, $\mathbf{F}_{enc}^{Last} \in \mathbb{R}^{S_1 \times H' \times W' \times D'}$, and another from the last stage of the decoder, $\mathbf{F}_{dec}^{Last} \in \mathbb{R}^{S_2 \times H \times W \times D}$. Here, $S_1$ (e.g., 512) is significantly larger than $S_2$ (e.g., 48), while $H'$, $W'$, and $D'$ are much smaller than $H$, $W$, and $D$, e.g., by a factor of 32. Since $\mathbf{F}_{enc}^{Last}$ contains high-level semantic information, we aim to combine it with the modality-conditioned text embedding for segmentation mask generation. To address the dimension mismatch, we first apply global average pooling over the spatial dimensions of $\mathbf{F}_{enc}^{Last}$, resulting in $\mathbf{e}_{vis} = \mathrm{GAP}(\mathbf{F}_{enc}^{Last})$ (see \Cref{fig: MulModSeg_arch}A). As $\mathbf{e}_{vis}$ is a vector, we can concatenate it with the modality-conditioned text embedding $\mathbf{e}_{txt}^{k}$ (see the text branch).

Our MulModSeg strategy with a 3D UNet or SwinUNETR backbone also performs a convolution operation with a kernel size of $1\times1\times1$ over the feature maps from the last stage of the decoder, \textit{i.e.}, $\mathbf{F}_{head}^{Pre} = \text{Conv-1}(\mathbf{F}{enc}^{Last})$. For example, the number of output channels for $\mathbf{F}_{head}^{Pre}$ can be set to 8 (smaller than $S_2$), therefore  $\mathbf{F}_{head}^{Pre} \in \mathbb{R}^{8\times H \times W \times D}.$ This convolution operation makes the subsequent mask prediction process more computationally efficient. The features are then processed by a text-driven segmentation module consisting of three sequential convolutional layers (Conv-2, see \Cref{fig: MulModSeg_arch}A) with $1 \times 1 \times 1$ kernels, and weights and biases parameterized by $\pmb{\theta}_{k}$ (see descriptions in text branch). The first two layers have 8 output channels, and the last layer has 1 output channel corresponding to the predicted class $([\text{CLS}]_k)$. The segmentation prediction for each class is computed as: 
$ \mathbf{P}_k = \text{Sigmoid}(((\mathbf{F}_{head}^{Pre} * \pmb{\theta}_k^1) * \pmb{\theta}_{k}^2) * \pmb{\theta}_{k}^3), $ where $\pmb{\theta}_k = \{\pmb{\theta}_k^1, \pmb{\theta}_{k}^2, \pmb{\theta}_{k}^3\}$ and $*$ for convolution. The predicted mask $\mathbf{P}_k \in \mathbb{R}^{1 \times H \times W \times D} $ denotes the foreground of each class in one \textit{vs.} all manner. By combining all different classes, we can obtain a multi-class prediction mask $\mathbf{\hat{Y}}$.

\noindent \textbf{Modality Driven Alternating Training} \label{sec:ALT}  
ALT shown in \Cref{fig: MulModSeg_arch}B is designed to enhance model training on multi-modal medical imaging data, specifically alternating between CT and MR modalities within each training iteration. The algorithm utilizes cyclic loaders for both CT and MR datasets, ensuring uninterrupted data feeding by looping back to the start once the end of a dataset is reached. During each iteration, the algorithm processes a batch from each modality: first CT, then MR by extracting images and labels, identifying the batch's modality, and feeding this information into the model to generate predictions. This methodical alternation between CT and MR batches allows for balanced exposure to both modalities, promoting model robustness and preventing bias towards either modality, thus enhancing the model's generalization capabilities across diverse medical imaging tasks. The overall ALT method is summarized in \Cref{alg: alt}.

\begin{algorithm}
\scriptsize
\caption{Alternating Training (ALT) with CT and MR Modalities}
\begin{algorithmic}[1]
\Procedure{TrainEpoch}{$CTLoader$, $MRLoader$, $model$, $optimizer$, $lossFunc$}
    \State $model.train()$
    \State $maxIter \gets max(len(CTLoader), len(MRLoader))$
    \State $cycleCT \gets itertools.cycle(CTLoader)$
    \State $cycleMR \gets  itertools.cycle(MRLoader)$
    \For{$iter \in range(maxIter)$}
        \State $batchCT \gets next(cycleCT)$
        \State $batchMR \gets next(cycleMR)$
        \For{$batch \in [batchCT, batchMR]$}
            \State $images, labels \gets batch['image'], batch['label']$
            \State $modality \gets batch['modality']$
            \State $logits \gets model(images, modality)$
            \State $loss \gets lossFunc(logits, labels)$
            \State $optimizer.zero\_grad()$
            \State $loss.backward()$
            \State $optimizer.step()$
        \EndFor
    \EndFor
\EndProcedure
\end{algorithmic}
\label{alg: alt}
\end{algorithm}

\section{Experiments}
\subsection{Experiments Setup}
We aim to validate the effectiveness of the proposed MulModSeg strategy in improving segmentation performance through a series of comprehensive experiments. To achieve this, several key research questions must be addressed. \textbf{\textit{Q1}}: How effective is text embedding in improving segmentation accuracy, and what type of embedding yields the best result? \textbf{\textit{Q2}}: Do modality-conditioned text embedding and alternating training (ALT) have a positive impact on the segmentation accuracy across different organ structures and imaging modalities (CT/MR) using various backbone architectures, specifically UNet and SwinUNETR? \textbf{\textit{Q3}}: Does MulModSeg outperform existing state-of-the-art (SOTA) methods in terms of performance?  \textbf{\textit{Q4}}: What is the impact of varying the ratio of CT to MR scans on the model's performance, especially in simulating more realistic, real-world scenarios? 

\noindent \textbf{Datasets}
For a fair comparison with existing methodologies, we assembled an unpaired multi-modal dataset for abdominal multi-organ segmentation, consisting of 162 CT scans and 54 MR scans from AMOS~\cite{ji2022amos} dataset. The focus is on segmenting 13 abdominal organs: spleen (SPL), right kidney (RKI), left kidney (LKI), gallbladder (GBL), esophagus (ESO), liver (LIV), stomach (STO), aorta (AOR), inferior vena cava (IVC), pancreas (PAN), right adrenal gland (RAD), left adrenal gland (LAG), and duodenum (DUO). Following established protocols, distinct preprocessing methods were applied to CT and MR scans to address modality discrepancies. CT scans are clipped to the intensity window $[-275, 125]$, and then normalized to $[0, 1]$, while MR scans were resampled to $[1.5 \times 1.5 \times 2.0]  ~ mm^3$, cropped to $96 \times 96 \times 96$ for training, and intensity histograms clipped by 0.5\% before min-max normalization to [0, 1]. Specifically, it means removing the lowest and highest 0.5\% of voxel values before normalization aiming at reducing the impact of outliers, like noise or extreme intensities, for more consistent training. The Multi-Modality Whole Heart Segmentation Challenge~\cite{zhuang2019evaluation} (MMWHS) dataset comprises 20 CT and 20 MR scans, collectively used for cardiac substructure segmentation, focusing on seven substructures: left ventricle (LV), right ventricle (RV), left atrium (LA), right atrium (RA), myocardium of LV (MY), ascending aorta (AA), and pulmonary artery (PA). Preprocessing involves resampling scans to $[1.5 \times 1.5 \times 2.0]  ~ mm^3$ resolution and normalizing intensities to $[0, 1]$ range, cropped to $96 \times 96 \times 96$ for training. 

\noindent \textbf{Balanced Data Splitting} 
We utilized an equal ratio of CT to MR scans for the AMOS dataset, each with 54 scans. Of these, 35 scans are used for training and 19 for testing. Similarly, the MMWHS dataset includes 20 scans per modality, split 75\% for training and 25\% for testing.

\noindent \textbf{Imbalanced Data Splitting} 
In the AMOS dataset, we implemented ratios of 2:1 and 3:1 for CT to MR scans, thus increasing the number of CT scans to deviate from the balanced splitting. For the MMWHS dataset, building on previous cross-modality segmentation research on this dataset~\cite{li2020towards,bastico2023simple}, we employed MR as the auxiliary modality and CT as the target modality. This choice was driven by the superior soft tissue contrast provided by MR, which offers more detailed information for segmenting heart substructures. We divided the CT data randomly and evenly to conduct a two-fold cross-validation. Each training iteration involved 20 MRs and 10 CTs, simulating a scenario of data scarcity for the target modality.

\label{sec:Imp}  
\noindent \textbf{Implementation Details} For the text branch, we utilized the pre-trained ``ViT-B/32" text encoder from CLIP as the deault setting. Since CLIP embeddings are derived from a fixed dictionary, we extracted and stored the text features in advance to reduce computational overhead during both training and inference. For the vision branch, we implemented both 3D UNet and SwinUNETR as backbones. Using SwinUNETR as an example, the vision encoder consists of four attention stages for encoding, each with two transformer blocks, and five CNN-based convolutional stages for decoding. In the encoder, a patch merging layer reduces spatial resolution by a factor of 2. Stage 1 begins with a linear embedding layer and transformer blocks that maintain the number of tokens at ${S_2} \times \frac{H}{2} \times \frac{W}{2} \times \frac{D}{2}$, with $S_2$ set to 48. The patch merging layer groups patches of size $2 \times 2 \times 2$, creating a $4 \times S_2$ feature embedding, followed by a linear layer that down-samples to $2 \times S_2$ of channels. This process repeats through stages 2, 3, and 4 \cite{tang_self-supervised_2022}. The five decoder stages together with skip connections gradually halve the channels and double the spatial resolution, with the final feature map, ${S_2} \times H \times W \times D$, corresponding to $F_{dec}^{Last}$ (\Cref{fig: MulModSeg_arch}A). The 3D UNet implementation follows a similar procedure, replacing the attention stages with additional convolutional stages in the encoder, using a different $S_2$ value of 64.

We utilized PyTorch 2.0 and MONAI 1.2~\cite{cardoso2022monai} to implement both our proposed MulModSeg strategy and other methods for comparison. All models were trained from scratch on a server with NVIDIA A100 GPUs. We employed the AdamW~\cite{loshchilov2017decoupled} optimizer with a warm-up cosine scheduler with an initial learning rate of $10^{-3}$ and a weight decay of $10^{-4}$ for 1000 epochs training including the first 10 epochs for warmup. The model of the last epoch is used for evaluation based on empirical experience. All hyperparameters are obtained through two-fold cross-validation over the training set unless otherwise specified. To avoid overfitting, on-the-fly data augmentation is applied, including random foreground and background patch sampling with a $1:1$ ratio and intensity shifting/scaling. A sum of Dice loss and Cross-entropy loss is used for training. For inference, an overlapping area ratio of 0.5 is applied via a sliding window strategy. 
\label{sec:evaluation_metrics}
The Dice score or Dice similarity coefficient is used as a metric for 3D segmentation results in the medical domain. It is defined as:
\begin{equation}
    \text{Dice}=\frac{2 \sum_{i=1}^I \mathbf{Y}_i \hat{\mathbf{Y}}_i}{\sum_{i=1}^I \mathbf{Y}_i+\sum_{i=1}^I \hat{\mathbf{Y}}_i},
\end{equation}
where $\mathbf{Y}_i$ and $\hat{\mathbf{Y}_i}$ denote the $i$-th ground truth and prediction of voxel values. $I$ represents the total number of the voxels. The Dice score ranges from 0 to 1, with 0 indicating no overlap and 1 representing a perfect match.

\begin{table}[h]
        \setlength{\tabcolsep}{2pt}
        \centering
        \resizebox{\columnwidth}{!}{
	\begin{tabular}{c|c|c|c} 
            \hline
		Embedding method & Prompt template & Avg. Dice$\uparrow$  (CT)&  Avg. Dice$\uparrow$ (MR)\\ \hline 
            Vision-Only & - & 82.50 & 81.91\\
            One Hot~\cite{zhang2021dodnet} & - & 86.64  & 84.50 \\
            BioBERT~\cite{lee2020biobert} & A \{CT/MR\} imaging of a [CLS]. & 86.62 & 84.59\\
            MedCLIP~\cite{wang2022medclip} & A \{CT/MR\} imaging of a [CLS]. & 86.30 & 84.10 \\ 
            V1-CLIP & A photo of [CLS]. & 85.60 & 84.30\\ 
            V2-CLIP  & There is [CLS] in this \{CT/MR\}. & 86.20 & 84.70\\ 
            V3-CLIP (\textbf{Ours}) & A \{CT/MR\} imaging of a [CLS].& \textbf{87.14} & \textbf{85.33}\\ 
            \hline
	\end{tabular}
        }
         \caption{Performance comparison of different text embeddings of MulModSeg on AMOS dataset with UNet backbone. V3-CLIP achieves the highest mean Dice scores for both CT and MR modalities. The \textbf{bolded} text represents the best performance. CT: computerized tomography, MR: magnetic resonance. The default Vison-Only model does not use text information and is trained with ALT. } 
	\label{tab: eff_text_emb}
\end{table}

\begin{table*}[h]
\centering
\resizebox{0.85\textwidth}{!}{%
\begin{tabular}{llcccccccccccccc}
\hline
\multicolumn{2}{c}{\multirow{2}{*}{Setting}} & \multicolumn{13}{c}{(\textbf{UNet}) Cat. Dice (\%) of Abdominal Organs $\uparrow$} & \multirow{2}{*}{Avg.$\uparrow$}\\
\cline{3-15}
& & SPL & RKI & LKI & GBL & ESO & LIV & STO & AOR & IVC & PAN & RAD & LAG & DUO \\
\hline
\multirow{2}{*}{w/o text}
& CT$\rightarrow$CT & 92.01  & 93.62 & 93.56 & 76.81 & 69.53 & 95.51 & 88.02  & 92.26  & 86.31 & 78.72 & 61.87 & 62.61 & 65.22 & 81.23 \\
& ALT$\rightarrow$CT & 92.90 & 94.19 & 94.31 & 77.02  & 71.26 & 95.62 & 88.87 & 93.38 & 86.86 & 80.91 & 63.82 & 64.95 & 68.42 & 82.50 \\
\hdashline
\multirow{1}{*}{\textbf{w/ text}}
&\textbf{ALT$\rightarrow$CT} & \textcolor{PineGreen}{94.18} & \textcolor{PineGreen}{95.62} & \textcolor{PineGreen}{95.53} &\textcolor{PineGreen}{84.57} &\textcolor{PineGreen}{78.56} &\textcolor{PineGreen}{96.30} & \textcolor{PineGreen}{91.97} &\textcolor{PineGreen}{94.82} & \textcolor{PineGreen}{89.07} &\textcolor{PineGreen}{86.65}& \textcolor{PineGreen}{70.02} & \textcolor{PineGreen}{76.60} & \textcolor{PineGreen}{78.97} & \textcolor{PineGreen}{87.14} \\

\midrule
\multirow{2}{*}{{w/o text}} 
& MR$\rightarrow$MR & 94.57 & 93.91 & 93.12 & 67.38 & 70.12 & 95.87 & 86.13 & 91.02  & 86.44 & 79.42 & 55.73 & 51.20 & 60.13 & 78.85 \\
& ALT$\rightarrow$MR & 95.20 & 95.58 & 95.32 & 70.71 & 74.42 & 96.37 & 88.18 & 91.26 & 87.82 & 81.01 & 60.83 & 63.82 & 64.28 & 81.91 \\

\hdashline
\multirow{1}{*}{\textbf{w/ text}} 
& \textbf{ALT$\rightarrow$MR} & \textcolor{blue}{96.06} & \textcolor{blue}{95.99} &\textcolor{blue}{95.85} & \textcolor{blue}{81.00} & \textcolor{blue}{78.55}  & \textcolor{blue}{97.40} & \textcolor{blue}{90.92} & \textcolor{blue}{92.28}  & \textcolor{blue}{90.20}   & \textcolor{blue}{87.03} & \textcolor{blue}{67.38} & \textcolor{blue}{66.32}  & \textcolor{blue}{70.30} & \textcolor{blue}{85.33} \\
\midrule
\addlinespace
\addlinespace
\midrule
\multicolumn{2}{c}{\multirow{2}{*}{Setting}} & \multicolumn{13}{c}{(\textbf{SwinUNETR}) Cat. Dice (\%) of Abdominal Organs $\uparrow$} & \multirow{2}{*}{Avg.$\uparrow$}\\
\cline{3-15}
& & SPL & RKI & LKI & GBL & ESO & LIV & STO & AOR & IVC & PAN & RAD & LAG & DUO \\
\midrule
\multirow{2}{*}{w/o text}
& CT$\rightarrow$CT & 94.02  & {95.39} & {95.22} & 82.41 &{77.57} & 95.89 & 88.76  & \textcolor{PineGreen}{94.50}  & 88.25 & 84.61 & 64.33 & 71.17 & 73.67 & 85.06 \\

& ALT$\rightarrow$CT & 94.49 & 95.03 & 95.11 & 79.39  &  \textcolor{PineGreen}{78.35} & 95.98 & 88.00 & 94.05 & {88.17 }& {88.17} & 62.95 & {72.44} & 73.18 & 85.02 \\

\hdashline
\multirow{1}{*}{\textbf{w/ text}}
& \textbf{ALT$\rightarrow$CT} & \textcolor{PineGreen}{95.32} & \textcolor{PineGreen}{95.75} & \textcolor{PineGreen}{95.43} 
&\textcolor{PineGreen}{84.24} & {77.56} &\textcolor{PineGreen}{96.49} & \textcolor{PineGreen}{90.04} 
&{94.05} & \textcolor{PineGreen}{88.42} &\textcolor{PineGreen}{84.91} & \textcolor{PineGreen}{69.03} & \textcolor{PineGreen}{72.92} & \textcolor{PineGreen}{75.81} & \textcolor{PineGreen}{86.15} \\

\midrule
\multirow{2}{*}{{w/o text}} 

& MR$\rightarrow$MR & 95.73 & 95.06 & 94.75 & \textcolor{blue}{69.94} & 75.30 & 96.24 & 84.58 & 90.35  & 87.34 & 81.71 & 56.97 & 57.47& 63.67 & 80.70 \\

& ALT$\rightarrow$MR & 95.79 & 95.81 &{95.42} & 67.03 & 76.24 & 96.28 & 86.19 & \textcolor{blue}{91.83} & 87.58 & 83.02  & 62.17 & 65.22 & 64.51 & 82.08 \\

\hdashline
\multirow{1}{*}{\textbf{w text}} 
& \textbf{ALT$\rightarrow$MR} & \textcolor{blue}{96.00} & \textcolor{blue}{95.84} & \textcolor{blue}{ 95.70} & {68.80} & \textcolor{blue}{76.30}  & \textcolor{blue}{96.57} & \textcolor{blue}{88.31} & {91.32}  & \textcolor{blue}{89.38}   & \textcolor{blue}{84.41} & \textcolor{blue}{65.52} & \textcolor{blue}{67.23}  & \textcolor{blue}{68.43} & \textcolor{blue}{83.37} \\

\hline

\end{tabular}
}%
\caption{ Dice scores on AMOS dataset with balanced data splitting of MulModSeg using UNet and SwinUNETR backbones. Results are shown for settings without text embedding (w/o text) and with text embedding (w/ text) across various training and testing scenarios. \textcolor{PineGreen}{Green} and \textcolor{blue}{blue} color represent the best performance for the CT and MR testing set, respectively.}
\label{tab: amos_dice}
\end{table*}

\begin{table*}[h]
\centering
\resizebox{1.0\textwidth}{!}{%
\begin{tabular}{llcccccccc}
\hline
\multicolumn{2}{c}{\multirow{2}{*}{Setting}}& \multirow{2}{*}{Avg.$\uparrow$} & \multicolumn{7}{c}{ Cat. Dice(\%) of Substructures of Heart (\textbf{UNet}) $\uparrow$} \\
\cline{4-10}
& & & MY & LA & LV & RA & RV & AA & PA \\
\hline
\multirow{2}{*}{w/o text}
& CT$\rightarrow$CT & 90.55 & 90.54 & 94.25 & 88.75 & 87.12 & 91.53 & 95.87 & 85.81 \\
& ALT$\rightarrow$CT & 90.41 & 90.80 & 94.56 & 88.82 & 85.37 & \textcolor{PineGreen}{92.00} & 95.10 & 86.24 \\

\hdashline
\multirow{1}{*}{\textbf{w text}} 
& \textbf{ALT$\rightarrow$CT} & \textcolor{PineGreen}{91.67} & \textcolor{PineGreen}{91.40} & \textcolor{PineGreen}{95.28} & \textcolor{PineGreen}{90.82} & \textcolor{PineGreen}{88.57} & {91.87} & \textcolor{PineGreen}{96.50} 
& \textcolor{PineGreen}{87.23}  \\
\midrule
\multirow{2}{*}{{w/o text}} 
& MR$\rightarrow$MR & 81.04 & 81.11 & 85.92 & 74.72 & 83.61 & 87.32 & 74.63 & 79.99 \\
& ALT$\rightarrow$MR & 82.82 & 81.00 & 87.03 & 79.03 & 85.62 & 88.06 & 76.90 & 82.22 \\

\hdashline
\multirow{1}{*}{\textbf{w text}} 
& \textbf{ALT$\rightarrow$MR} & \textcolor{blue}{85.15} & \textcolor{blue}{83.55} & \textcolor{blue}{89.29} & \textcolor{blue}{83.27} 
& \textcolor{blue}{87.45} & \textcolor{blue}{88.30} & \textcolor{blue}{80.52} 
& \textcolor{blue}{83.67} \\
\hline
\end{tabular}

\hspace{0.25cm}
\begin{tabular}{cccccccc}
\hline
\multirow{2}{*}{Avg.$\uparrow$} &\multicolumn{7}{c}{ Cat. Dice(\%) of Substructures of Heart (\textbf{SwinUNETR}) $\uparrow$} \\
\cline{2-8}
& MY & LA & LV & RA & RV & AA & PA\\
\hline
90.87 & 90.53 & 95.01 & \textcolor{PineGreen}{90.49} & 86.42 & 90.89 & 95.95 & 86.79 \\
91.11 & 90.98 & 94.84 & 90.26 & 87.35 & \textcolor{PineGreen}{92.22} & 96.07 & 86.27 \\

\hdashline
\textcolor{PineGreen}{91.44} & \textcolor{PineGreen}{91.10} & \textcolor{PineGreen}{95.13} & 90.25 & \textcolor{PineGreen}{88.00} & 91.89 
& \textcolor{PineGreen}{96.33} & \textcolor{PineGreen}{87.36} \\
\midrule
83.07 & 83.17 & \textcolor{blue}{90.39} & 81.05 & 84.44 & 87.67 & 76.41 & 78.36 \\
83.33 & 83.07 & 88.43 & 81.52 & \textcolor{blue}{85.88} & 86.79 & 76.67 & \textcolor{blue}{80.94} \\

\hdashline
\textcolor{blue}{83.85} & 82.48 & 90.22 & \textcolor{blue}{82.49} & 85.32 & \textcolor{blue}{88.39} 
& \textcolor{blue}{77.24} & 80.79 \\
\hline
\end{tabular}
}%
\caption{Dice scores with UNet and SwinUNETR backbone for MMWHS dataset with balanced data splitting. \textcolor{PineGreen}{Green} and \textcolor{blue}{blue} color represent the best performance for the CT and MR testing set, respectively.}
\label{tab: mmwhs_dice}
\end{table*}

\subsection{Results and Discussion}

\noindent \textbf{Effectiveness of Different Text Embeddings (\textbf{\textit{Q1}})}
To investigate the impact of integrating text embeddings with vision-only models and their variants on segmentation performance, we conducted a comprehensive series of studies using different text embeddings: One Hot, BioBERT, MedCLIP, and our proposed V1, V2, and V3-CLIP models. The results, presented in  \Cref{tab: eff_text_emb}, highlight the performance on AMOS dataset with a UNet backbone. Our experiments show a significant advantage of using modality-conditioned text embeddings over conventional methods. Specifically, the V3-CLIP embedding achieved the highest mean Dice scores of 87.14 for CT and 85.33 for MR, demonstrating its superior capability in enhancing feature representation and segmentation accuracy. \textit{This improvement underscores the importance of leveraging modality-specific textual information to enrich the context and details captured by the segmentation model}.

A closer examination of the results reveals that the Vision-Only model, which does not utilize any text information, had the lowest average Dice scores, with 82.50 for CT and 81.91 for MR, highlighting the limitations of purely vision-based models. The One Hot encoding method showed slight improvements but still lacked semantic depth, achieving 86.64 for CT and 84.50 for MR. BioBERT and MedCLIP, leveraging pre-trained language representations, provided better performance by capturing medical terminology and context. 

Although various text embedding methods significantly improve performance over the baseline vision-only model (see \Cref{tab: eff_text_emb}), the choice of medical prompt template is also crucial. The performance differences between V1-CLIP, V2-CLIP, and V3-CLIP stem from the specificity of their prompt templates. V1-CLIP uses a generic prompt (``A photo of [CLS]"), which lacks medical context. V2-CLIP improves slightly by mentioning the modality but remains somewhat vague (``There is [CLS] in this {CT/MR}"). V3-CLIP (Ours), with its prompt focused on ``A {CT/MR} imaging of [CLS]," is the most aligned with the medical domain, leading to better semantic conditioning and, thus, superior performance in segmentation tasks. Consequently, V3-CLIP is utilized as the default setting. However, selecting an optimal template remains an open challenge for medical image-text-vision models, and we encourage future research to explore this area further.

\begin{figure}[h]
	\centering
	\includegraphics[width=0.475\textwidth]{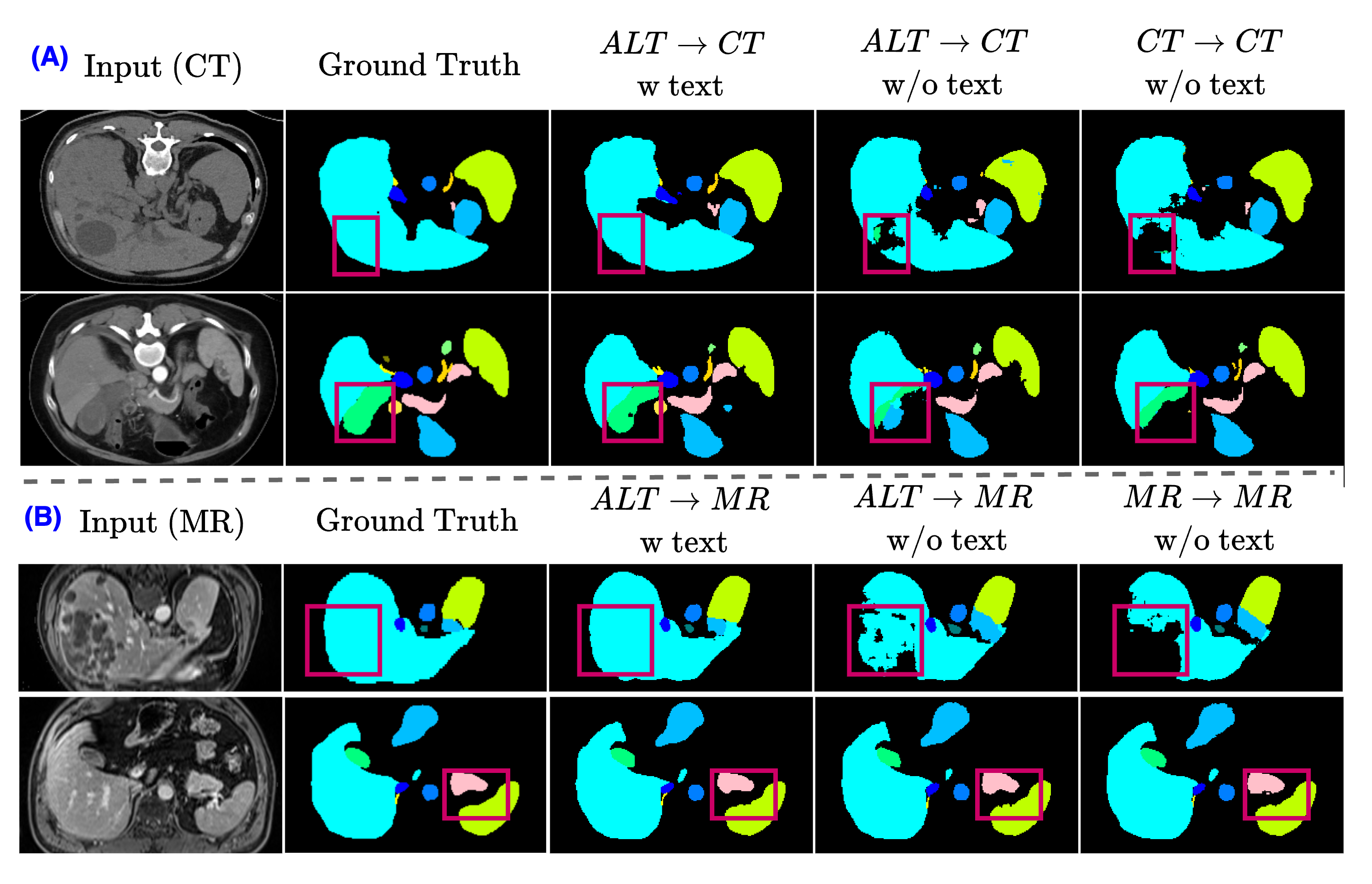}
	\caption{Visual comparison of segmentation results on AMOS dataset using the UNet backbone. {\color{red}Red boxes} highlight areas where MulModSeg demonstrates improved predicted details compared to baselines.} 
        \label{fig: s1_amos_qualitative}
\end{figure}

\begin{table*}[h]
\centering
\scriptsize
\resizebox{0.65\textwidth}{!}{
\begin{tabular}{l|c|c|c|c|c|c|c|c}
\hline
 \multirow{2}{*}{Model} & \multirow{2}{*}{Avg. $\uparrow$} & \multicolumn{7}{c}{Dice of Substructures of Heart (UNet) $\uparrow$}\\
 \cline{3-9}
 & & MY & LA & LV & RA & RV & AA & PA\\
 \hline
 Baseline~\cite{li2020towards} & 87.06 & 87.02 & 89.22 & 90.86 & 83.86 & 84.60 & 92.52 & 81.34 \\
 Fine-tune~\cite{li2020towards} & 87.69 & 87.16 & 90.40 & 90.79 & 84.43 & 85.26 & 92.74 & 83.05 \\
 Joint-training\cite{li2020towards} & 87.43 & 86.65 & 90.76 & 91.23 & 82.78 & 84.92 & 93.02 & 82.66 \\
 X-shape \cite{valindria2018multi} & 87.67 & 87.19 & 89.79 & 90.94 & 85.51 & 84.44 & 93.43 & 82.40 \\
 Zhang \textit{\etal}\cite{zhang2018translating} & 88.50 & 87.81 & 91.12 & 91.34 & 85.14 & 86.31 & 94.30 & 83.42 \\
 Li \textit{\etal}\cite{li2020towards} & 90.12 & 89.34 & 91.90 & 92.67 & 87.47 & 88.14 & \textbf{95.95} & 85.38 \\
 Bastico \textit{\etal}\cite{bastico2023simple} & 90.77 & 90.06 & 92.68  & \textbf{93.77} & 88.22 & 90.85 & 94.70 & 84.52\\
 \textbf{Ours} & \textbf{92.72} & \textbf{92.32} &\textbf{ 92.73 }& 89.85 & \textbf{93.42} & \textbf{94.80} & 95.48 & \textbf{90.46}\\
\hline
\hline

\hline
\multirow{2}{*}{Model} & \multirow{2}{*}{Avg. $\uparrow$} & \multicolumn{7}{c}{Dice of Substructures of Heart (SwinUNETR) $\uparrow$}\\
 \cline{3-9}
 & & MY & LA & LV & RA & RV & AA & PA\\
 \hline
 Baseline\cite{zhang2018translating}  & 85.32 & 84.79 & 88.12 & 89.80 & 81.04 & 84.08 & 77.58 & 76.36\\
 Fine-tune\cite{zhang2018translating}  & 86.09 & 82.57 & 90.03 & 87.92 & 83.29 & 85.51 & 90.28 & 83.06\\
 Joint-training\cite{zhang2018translating} & 87.99 & 86.25 & 92.07 & 91.97 & 85.35 & 87.98 & 89.55 & 82.76\\
Bastico \textit{\etal}\cite{bastico2023simple}  & 89.33 & 88.13 & 91.64 & \textbf{92.39} & 86.36 & 89.33 & 93.39 & 84.05\\
  \textbf{Ours} & \textbf{93.31} & \textbf{93.15} & \textbf{93.50 }& 91.59 & \textbf{92.20} & \textbf{95.63} & \textbf{95.83} & \textbf{91.27}\\
\hline

\end{tabular}}
\caption{Quantitative comparison with other methods for cross-modality medical image segmentation on the target modality (CT). All the techniques have the same UNet \cite{valindria2018multi} and SwinUNETR \cite{hatamizadeh2021swin} baseline, are trained using 20 MRs and 10 CTs and are evaluated on the test set of MMWHS dataset. The mean Dice score is reported, as well as the ones for all the heart substructures.}
\label{tab: comp_sota_MMWHS}
\end{table*}

\noindent \textbf{Positive Impacts of Modality-Conditioned Text Embedding and ALT (\textbf{\textit{Q2}})}
\Cref{tab: amos_dice} presents Dice scores for unpaired multi-modal abdominal multi-organ segmentation using UNet and SwinUNETR backbones on the AMOS dataset. It compares segmentation accuracy across various settings: without text embedding (w/o text) and with text embedding (w text), and for different scenarios including CT (for training) $\rightarrow$ CT (for testing), ALT $\rightarrow$ CT, MR $\rightarrow$ MR, and ALT $\rightarrow$ MR. For both UNet and SwinUNETR backbones, configurations with text embeddings show substantial improvements over those without. For instance, with the UNet backbone, the ALT + Text setup yields an average Dice score of 87.14 on the CT test set, compared to 82.50 without text embeddings, and an increase from 81.91 to 85.33 on the MR test set. The SwinUNETR backbone shows similar trends, further validating our approach. Additionally, the ALT procedure alone outperforms setups without ALT, highlighting its effectiveness.  Similar enhancements of the details in predicted masks can be observed in visual comparison results (as shown in~\Cref{fig: s1_amos_qualitative}). 

\Cref{tab: mmwhs_dice} illustrates the significant performance gains achieved by the MulModSeg framework using modality-conditioned text embeddings and the ALT procedure for unpaired multi-modal cardiac substructure segmentation within MMWHS dataset. For the UNet backbone, the ALT + Text configuration achieves an average Dice score of 91.67 on the CT test set, compared to 90.41 without text embeddings, and an increase from 82.82 to 85.15 on the MR test set. Similar trends are observed with the SwinUNETR backbone. Additionally, the ALT procedure alone shows improved performance over non-ALT configurations. \textit{These results confirm that integrating text embeddings with ALT significantly enhances the model's generalization and segmentation accuracy across modalities. This demonstrates the robustness and effectiveness of the MulModSeg strategy in diverse cardiac and abdominal multi-organ segmentation tasks.}

\begin{table*}[h]
\centering
\resizebox{0.8\textwidth}{!}{%
\begin{tabular}{llcccccccccccccc}
\hline
\multicolumn{2}{c}{\multirow{2}{*}{Ratio of CT:MR}} & \multicolumn{13}{c}{AMOS-CT Testing Cat. Dice $\uparrow$} & \multirow{2}{*}{Avg.$\uparrow$}\\
\cline{3-15}
& & SPL & RKI & LKI & GBL & ESO & LIV & STO & AOR & IVC & PAN & RAD & LAG & DUO \\
\hline

\multirow{4}{*}{MulModSeg}

& 1:1 &{94.18} & {95.62} & {95.53} &{84.57} &{78.56} &{96.30} & {91.97} &{94.82} &{89.07} &{86.65}& {70.02} & {76.60} & {78.97} & {87.14} \\
& 2:1 & 95.23 & 95.72 & 96.01 & 85.02  & 80.07 & 96.40 & 92.04 & 95.07 & 89.91 & 86.65 & 71.93 & 76.75 & 80.80 & 87.82 \\
& 3:1 & 95.72 & 96.14 & 96.22 & 86.69 & 83.85 & 97.22 & 93.77  & 95.39 & 90.29 & 87.99 & 72.48 & 79.30 & 80.89 & \textbf{88.92}\\

\hline
\addlinespace
\hline
\multicolumn{2}{c}{\multirow{2}{*}{Ratio of CT:MR}} & \multicolumn{13}{c}{ AMOS-MR Testing Cat. Dice $\uparrow$} & \multirow{2}{*}{Avg.$\uparrow$}\\
\cline{3-15}
& & SPL & RKI & LKI & GBL & ESO & LIV & STO & AOR & IVC & PAN & RAD & LAG & DUO \\
\hline
\multirow{4}{*}{MulModSeg}
& 1:1 & {96.06} & {95.99} &{95.85} & {81.00} &{78.55}  &{97.40} & {90.92} & {92.28}  & {90.20}   & {87.03} & {67.38} &{66.32}  & {70.30} & {85.33} \\
& 2:1 & 96.42 & 95.94 & 95.81 & 76.11  & 79.30 & 97.50 & 90.93 & 92.54 & 90.83 & 85.84 & 64.24 & 72.01 & 72.53 & 85.38 \\
& 3:1 & 96.40 & 96.06 & 96.18 & 77.14 & 79.56 & 97.58 & 91.17  & 92.15 & 90.59 & 86.62 & 64.28 & 71.42 & 72.28 & \textbf{85.49}\\
\hline

\end{tabular}}
\caption{Dice scores for abdominal multi-organ segmentation on AMOS dataset using the UNet backbone, with varying CT to MR scan ratios (1:1, 2:1, and 3:1). Results show the impact of different data ratios on segmentation performance.}
\label{tab: vary_ct_mr_ratio}
\end{table*}

\noindent \textbf{Comparison with State-of-the-Art Methods (\textbf{\textit{Q3}})}
In benchmarking the performance of our MulModSeg framework against several state-of-the-art methods for cross-modality medical image segmentation on the MMWHS dataset, we observed significant improvements across most of the heart substructures. As shown in \Cref{tab: comp_sota_MMWHS}, our method was evaluated using both the UNet and SwinUNETR backbones, with an emphasis on the target modality (CT). MulModSeg achieved the highest average Dice scores of 92.72 for the UNet backbone and 93.31 for the SwinUNETR backbone. These results underscore the effectiveness of our approach in leveraging modality-conditioned text embeddings and the ALT procedure. Specifically, our framework demonstrated remarkable improvements in segmenting critical cardiac substructures such as the MY, RA, and RV. For instance, with the UNet backbone, MulModSeg achieved Dice scores of 92.32 for MY, 93.42 for RA, and 94.80 for RV, surpassing previous methods by a considerable margin. Similarly, with the SwinUNETR backbone, our method attained Dice scores of 93.15 for MY, 92.20 for RA, and 95.63 for RV, indicating consistent performance enhancements across different architectural setups. \textit{These results highlight the robustness and superior performance of our method, attributed to the innovative use of modality-conditioned text embeddings and the ALT procedure, which enhance the model's ability to leverage information from unpaired multi-modal datasets.}

\noindent \textbf{Impact of Varying CT to MR Scan Ratios (\textbf{\textit{Q4}})}
We also explored the impact of varying the ratio of CT to MR scans on the segmentation performance to simulate more realistic clinical scenarios. \Cref{tab: vary_ct_mr_ratio} presents the Dice scores for different CT:MR ratios (1:1, 2:1, and 3:1) on AMOS dataset using the UNet backbone. The results indicate that increasing the number of CT scans relative to MR scans generally enhances segmentation performance. For instance, the average Dice score for the 3:1 CT:MR ratio was 88.92, compared to 87.14 for the 1:1 ratio. \textit{This trend is consistent across both CT and MR testing sets, suggesting that our approach can effectively leverage the availability of more CT data to improve segmentation accuracy while maintaining robust performance across modalities.}

\noindent \textbf{Ablation Study} \Cref{tab: tab_ab2} demonstrates the significant impact of using modality-conditioned text embeddings and the ALT training procedure on segmentation performance. The configuration combining text embeddings with ALT training achieved the highest average Dice scores of 87.14 for CT and 85.33 for MR, showcasing the superior capability of this approach. In contrast, models trained without ALT or using only CT or MR images exhibited substantially lower Dice scores, underscoring the necessity of modality-specific textual information and the ALT procedure for optimal performance. Specifically, the model trained with CT data and text embeddings achieved a Dice score of 86.40 for CT but only 69.72 for MR, while the model trained with MR data and text embeddings showed a significant drop in CT performance, illustrating the challenges of cross-modality training without the ALT method. 

To further evaluate the impact of medical text prompts, we tested with confused prompts. \Cref{tab: tab_ab_mis} shows the effect of using mistaken prompts (CT prompt for MR image, MR prompt for CT image) on the average Dice scores for CT and MR images in the AMOS test set with a UNet backbone. Without mistakes, the model achieves higher Dice scores: 87.14 for CT and 85.33 for MR. With mistaken prompts, the scores drop to 86.37 for CT and 85.10 for MR.

In \Cref{tab: tab_ab3}, we show the number of network parameters, FLOPs, and inference time for different settings on the AMOS dataset. FLOPs and inference time are calculated with a $96 \times 96 \times 96$ input using sliding window approaches. MulModSeg (with text embedding) has comparable parameters and inference time to baselines without text embedding. Interestingly, MulModSeg slightly reduces FLOPs, likely due to the $1\times1\times1$ Conv-1 and Conv-2 operations, which reduce intermediate feature map channels.

\begin{table}[h]
    \scriptsize
     \setlength{\tabcolsep}{4pt}
        \centering
	\begin{tabular}{cc|c|c} \hline
		Text Emb. & Training & Avg. Dice$\uparrow$ (CT)& Avg. Dice$\uparrow$ (MR)\\ \hline
            $\bullet$& $ALT$ & \textbf{87.14} & \textbf{85.33} \\ 
            $\circ$ & $ALT$ & 82.50 & 81.91\\ 
            $\bullet$ & $CT$ & 86.40 & 69.72\\ 
            $\bullet$ & $MR$ & 60.53 & 82.70\\ 
            $\circ$ & $CT$ & 81.23 & 55.45\\ 
            $\circ$ & $MR$ & 46.64 & 78.85\\ 
            \hline
	\end{tabular}
        \caption{ Ablation study for the MulModSeg strategy.  $\bullet$: with, $\circ$: without. The \textbf{bolded} font represents the best performance in AMOS testing sets with UNet backbone.}
	\label{tab: tab_ab2}
\end{table}

\begin{table}[h]
     \scriptsize
     \setlength{\tabcolsep}{4pt}
        \centering
	\begin{tabular}{c|c|c} \hline
		Mistaken Prompts & Avg. Dice$\uparrow$ (CT)& Avg. Dice$\uparrow$ (MR)\\ \hline
            No & \textbf{87.14} & \textbf{85.33} \\ 
            Yes & 86.37 & 85.10\\ 
            \hline
	\end{tabular}
    \caption{During testing, CT prompts were mistakenly applied to MRI images, and MR prompts to CT images. The \textbf{bolded} font indicates the best performance on AMOS test sets with the UNet backbone.}
 
    \label{tab: tab_ab_mis}
\end{table}

\begin{table}[h]
     \scriptsize
        \setlength{\tabcolsep}{2pt}
        \centering
	
        \begin{tabular}{cc|c|c|c} \hline
		Text Emb. & Backbone & Params. (M) & FLOPs (G) & Time (s) \\ \hline
            $\bullet$ & UNet & 19.4 & 1002.20 & 3.272 \\ 
            $\circ$  & UNet & 19.1 & 1002.54 & 2.402\\ 
           $\bullet$ & SwinUNETR & 62.6 & 329.59 & 4.197\\
           $\circ$  & SwinUNETR & 62.2 & 329.84 & 3.387\\ 
            \hline
	\end{tabular}
        \caption{Comparison of the number of parameters, FLOPs, and averaged inference time for with/without using text embedding in AMOS CT testing experiments. $\bullet$: with, $\circ$: without.} 
	\label{tab: tab_ab3}
\end{table}

\section{Conclusion}

In this work, we introduced MulModSeg, a multi-modal segmentation strategy enhancing CT and MR medical image segmentation. MulModSeg leverages modality-conditioned text embeddings and an alternating training (ALT) procedure, integrating modality-specific information into existing encoder-decoder frameworks without significant architectural changes. Extensive experiments showed that MulModSeg significantly improves segmentation accuracy and robustness over state-of-the-art methods, achieving higher Dice scores for abdominal multi-organ and cardiac substructure segmentation tasks. The method's adaptability to different imaging modalities and balanced/imbalanced training scenarios across unpaired datasets ensure practical clinical application. 

\noindent{\textbf{Limitation}}  
While our proposed MulModSeg strategy significantly enhances segmentation accuracy across multi-modal CT/MR imaging using a single model, future work will focus on expanding its application to additional imaging modalities, such as X-ray and ultrasound, as well as broader clinical scenarios.


{\small
\bibliographystyle{ieee_fullname}
\bibliography{egbib}
}

\end{document}